\documentclass[preprint,12pt, a4paper]{elsarticle}

\usepackage{amssymb}
\usepackage{hyperref}

\journal{Science of Computer Programming}

\begin{document}

\begin{frontmatter}

\title{VeGAn-Tool: A Fuzzy-logic Approach for Value-based Goal Model Analysis}

\author{C. Cano-Genoves}
\author{E. Insfran}
\author{S. Abrahão}

\address{IUMTI, Universitat Politècnica de València, Valencia, Spain}

\begin{abstract}

Goal-oriented analysis tools are used to assess goal models and assist analysts in decision-making. We introduce the VeGAn-Tool, which prioritizes goals according to their qualitative importance for the stakeholders and propagates this information in the goal model according to the different types of relationships. The FTOPSIS technique is used to calculate the value of each intentional element by employing the fuzzified importance (importance level fuzzified and refined by a confidence level) and the impact among the related intentional elements. The result is a prioritized goal model according to the value of each intentional element from the stakeholders’ point of view.

\end{abstract}


\end{frontmatter}


\section*{Metadata}

\begin{table}[!h]
\footnotesize
\begin{tabular}{|l|p{6.5cm}|p{6.5cm}|}
\hline
\textbf{Nr.} & \textbf{Code metadata description} & \textbf{} \\
\hline
C1 & Current code version & v2.0.1 \\
\hline
C2 & Permanent link to code/repository used for this code version & \url{https://github.com/CarlosCanoGenoves/VeGAn-Tool} \\
\hline
C3  & Permanent link to Reproducible Capsule & \url{https://doi.org/10.5281/zenodo.8142696} \\
\hline
C4 & Legal Code License & Eclipse Public License (EPL-2.0) \\
\hline
C5 & Code versioning system used & git \\
\hline
C6 & Software code languages, tools, and services used & Java, Eclipse Modeling Tools, Eclipse OCL, Jackson, JGoodies \\
\hline
C7 & Compilation requirements, operating environments and dependencies & Java JDK 15.0.2 or higher, Eclipse Modeling Tools \\
\hline
C8 & If available, link to developer documentation/manual & 
\\
\hline
C9 & Support email for questions & carcage1@dsic.upv.es \\
\hline
\end{tabular}
\caption{Code metadata}
\end{table}

\section{Motivation and significance}

Stakeholders’ goals and intentions with respect to the system to be developed can be represented using goal models. These models are typically employed during early requirements elicitation, as they help understand the motivations underlying the system requirements and thus help the system to be developed in accordance with its stakeholders’ interests.

Goal models are usually analyzed by means of satisfaction, since this makes it possible to reason about possible conflicts among them and determine whether particular conditions can eventually lead to the satisfaction of a set of goals by considering different alternatives. Tools such as OpenOME \cite{Interactive} and jUCMNav \cite{GRLEvaluation} follow this approach, and jUCMNav also allows the assignment of an importance level (qualitative or quantitative) with which to supplement the information regarding the satisfaction results in order to help analysts choose among alternative goals.

The VeGAn-Tool (Value-based Goal-oriented Analysis) \cite{MODELS} automates an approach that analyzes goal models through the concept of value, on the basis of a relative qualitative importance and a confidence level so as to identify which goals are most valuable to the stakeholders. The approach also considers the relationships among intentional elements in the whole goal model. As single importance and confidence levels are assigned to each intentional element, some consensus is required among representatives of the stakeholder if there are more than one. The main feature of the VeGAn-Tool is the prioritization of goals according to the stakeholders’ preferences (relative importance), which are weighted and propagated according to the different types of relationships in the goal model. A variation of the FTOPSIS technique (Fuzzy Technique of Order Preference Similarity to the Ideal Solution) is used to calculate the value of each intentional element by employing the fuzzified importance (importance level fuzzified and refined by the confidence level) and the impact among the related intentional elements.

We consider that the VeGAn-Tool could complement other satisfaction-based goal analysis tools, as analysts can focus on determining the satisfiability of only those goals that are most valuable to the stakeholders, thus reducing the alternatives for the decision-making process.

In order to use the tool, a goal model must first be loaded. The model must be stored in an XMI file according to the Ecore metamodel provided along with the tool, although it is also possible to import a model from the piStar tool [4]. Optionally, a picture of the goal model can also be loaded in order to be able to view it during the prioritization and analysis.

Next, each stakeholder must prioritize each of their intentional elements in the goal model by assigning both a level of importance and a level of confidence. An analyst (usually the project leader) must then prioritize the stakeholders, since they may have different levels of importance in the project. The importance level represents how important each intentional element is, while the confidence level represents how confident the stakeholder is in the importance assigned to that intentional element.

Finally, it is necessary to press the propagation button, which performs the calculations required to determine the value for each intentional element. These calculations provide a \textit{local value}, which represents the value of the intentional element for the stakeholder considering only what the intentional elements in the scope of the stakeholder are, and a \textit{global value}, which represents the value of the intentional element for the stakeholder but considering the whole goal model. The information concerning the related intentional elements that positively or negatively contribute to the corresponding value is provided for traceability and explainability purposes.

Each time the value calculation is performed, the priorities are modified and the result is then versioned and stored in an XMI file, thus making it possible to analyze the evolution of priorities in time or compare alternative prioritizations in order to choose that which better fits the understanding of the system to be developed.

\section{Tool description}

The tool has been developed in Java using Eclipse Modeling Tools (EMT), Eclipse OCL plugin, Jackson and JGoodies. EMT has been used to work with models and generate the code of the model automatically. The Eclipse OCL plugin has been employed in order to validate the model and generate derived attributes. The Jackson library has been used to work with JSON files, which are the output of the piStar tool. The JGoodies library has been used to make the interface more visually pleasing and enjoyable.

\subsection{Tool architecture}

The tool architecture is divided into four packages: \textit{goalModel}, \textit{VEGAN}, \textit{piStar} and \textit{VISUAL}.
The \textit{goalModel} package corresponds to the goal model structure, which has been automatically generated from an Ecore, while the \textit{VEGAN} package is the cornerstone of the tool as it is responsible for the analysis. The \textit{piStar} package is in charge of importing models from the piStar tool, and the \textit{VISUAL} package is in charge of the user interface of the tool.

\subsection{Tool functionalities}

The main functionality of the tool is the analysis of goal models by means of value, for which it calculates both a local and a global value. The remaining functionalities are derived from this. For example: i) information regarding where the value of the goal comes from; ii) sorting goals by value and iii) storing the evolution of importance and value.

\section{Illustrative example}

\begin{figure*}[h]
  \centering
  \includegraphics[scale=0.12]{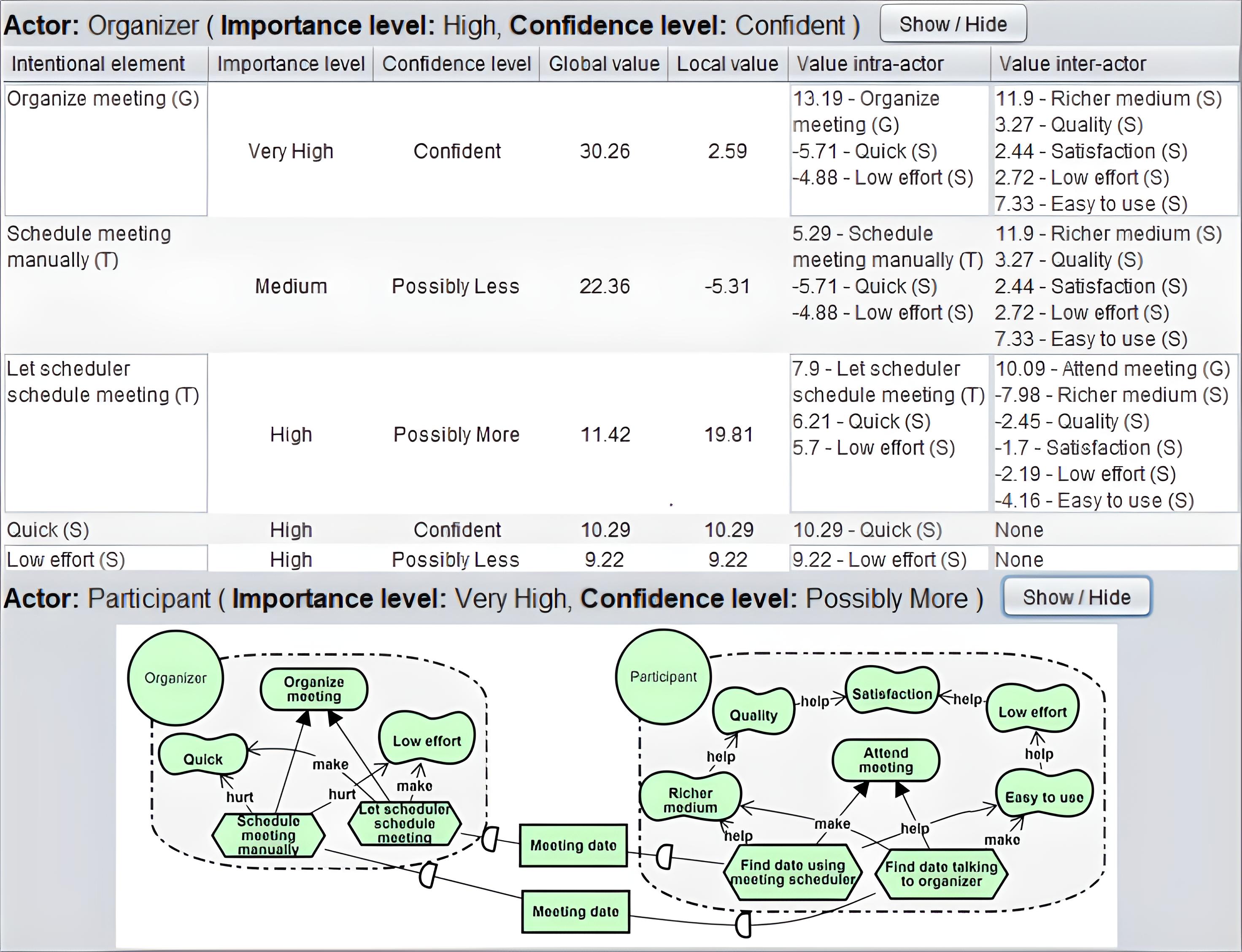}
  \caption{Result of the analysis of the goal model of \textit{Meeting Scheduler} system}
  \label{fig1}
\end{figure*}

Figure~\ref{fig1} shows an example of the use of the tool taken from \cite{MODELS}. The table at the top of the figure shows the analysis of the goal model in the picture at the bottom of the figure. The first column indicates the name of the goal analyzed, while columns two and three respectively indicate the importance and confidence of the importance assigned to each goal; columns four and five respectively show the \textit{global value} (the value it has for all stakeholders) and the \textit{local value} (the value it has for the interested stakeholder) calculated by the analysis, the value is a number between -100 (worthless) and 100 (valuable);
column six shows the value originating from intentional elements of the same actor (how the local value is calculated), and column seven shows the value originating from intentional elements of other actors (columns 6 and 7 together correspond to the global value). A video demonstration of the VeGAn-Tool is available at \url{https://youtu.be/b-6zQ63vSa4}.

\section{Impact}

From the research point of view, the proposed tool combines fuzzy-logic and multi-criteria decision–making in order to deal with the prioritization of intentional elements by different stakeholders, considering the uncertainty (confidence) of the relative importance. It also provides a systematic propagation mechanism with which to perform the analysis of the impact of related intentional elements according to the nature of the types of relationships among them. This is a complex and error-prone task to be performed manually, particularly in the case of large goal models.

Fuzzy logic prioritization addresses issues through the use of classical quantitative and qualitative techniques \cite{Comparison}. The difficulty of the former lies in the need to assign specific single values from a large set of options (e.g., 0..100). However, the latter are easier to use owing to the small number of options (e.g., low, med, high) but have the problem of a lack of accuracy. We have combined both techniques using qualitative prioritization, which are internally converted into fuzzy numbers (a range of possible values) in order to quantitatively perform the systematic propagation and decide which alternatives to choose.

We performed an experiment \cite{MODELSWARD} in order to compare our tool results (value analysis) with those obtained with a tool that implements the satisfaction analysis with quantitative importance (satisfaction analysis). The initial prioritizations of intentional elements were similar with both tools. However, the participants in the experiment were more satisfied with the results obtained from the value analysis than with those obtained from the satisfaction analysis. However, further experimentation is needed in order to confirm this finding.

From the point of view of industry, the VeGAn-Tool allows practitioners to use goal models following the principles of Value-Based Software Engineering (VBSE) \cite{VBSE}. For instance, software developers that follow incremental development approaches may need to identify which goals are most valuable for the stakeholders so as to include them as part of the increments to be delivered first.

\section{Conclusions}

We have presented VeGAn-Tool, a goal-oriented analysis tool that prioritizes individual intentional elements according to the relative importance indicated by the stakeholders. The tool uses fuzzy-logic and multi-criteria decision-making to systematically propagate this information throughout the different types of relationships of the goal model using FTOPSIS, and this has several benefits when compared to classical qualitative/quantitative approaches. The result is a prioritized goal model with value that facilitates the alignment of the software system development activities with the stakeholders’ and the organizational goals.

\section{Future Plans}

We plan to improve the interoperability of the VeGAn tool with other goal modeling tools (e.g., jUCMNav, OpenOME) allowing to import their goal models and their corresponding graphical representation. In addition, studying means to resize or move the intentional elements. We shall also study explainability techniques with which to improve the understanding of how the values of the intentional elements are calculated. Versioning of prioritization and calculated values is currently stored, but a comparison of versions is under development. Finally, we also plan to conduct experiments with end users in order to collect more evidence regarding the effectiveness of our tool and improve its usability.

\section*{Acknowledgements}

This work was supported by the grant TIN2017-84550-R (\url{Adapt@Cloud} project) funded by \url{MCIN/AEI/10.13039/501100011033} and the “Programa de Ayudas de Investigación y Desarrollo” (\url{PAID-01-17}) from the Universitat Politècnica de València.

\newpage
\bibliographystyle{elsarticle-num}
\bibliography{references}

\end{document}